\documentclass[a4paper,fleqn]{cas-dc}

\usepackage[authoryear]{natbib}
\usepackage{eurosym}
\usepackage{hyperref}
\usepackage[super]{nth}
\usepackage{soul}

\usepackage[acronym]{glossaries}
\newglossarystyle{myglossarystyle}
{%
  {% start of glossary
   \noindent
    \begin{tabular}{p{0.07\textwidth}p{0.36\textwidth}}
    \toprule
  }%
  {% end of glossary
     \bottomrule
     \end{tabular}%
  }%
  \renewcommand*{\glsgroupheading}[1]{}%
}
\makeglossaries
\newacronym{cheops}{CHEOPS}{CHaracterising ExOPlanet Satellite}
\newacronym{esa}{ESA}{European Space Agency}
\newacronym{soc}{SOC}{Science Operations Center}
\newacronym{moc}{MOC}{Mission Operations Center}
\newacronym{inta}{INTA}{Instituto Nacional de Técnica Aeroespacial}
\newacronym{ecss}{ECSS}{European Cooperation for Space Standardization}
\newacronym{or}{OR}{Observation Request}
\newacronym{pht2}{PHT2}{Proposal Handling Tool phase 2}
\newacronym{fds}{FDS}{Flight Dynamics System}
\newacronym{gs}{G/S}{Ground Station}
\newacronym{hmi}{HMI}{Human-machine interface}
\newacronym{sw}{SW}{Software}
\newacronym{ceit}{CEIT}{Centro Espacial INTA Torrejón}
\newacronym{mps}{MPS}{Mission Planning System}
\newacronym{gfts}{GFTS}{Generic File Transfer System}
\newacronym{mcs}{MCS}{Mission Control System}
\newacronym{parc}{PARC}{MCS Packet Archive}
\newacronym{obt}{OBT}{On-Board Time}
\newacronym{utc}{UTC}{Coordinated Universal Time}
\newacronym{nasa}{NASA}{National Aeronautics and Space Administration}
\newacronym{must}{MUST}{Mission Utility and Support Tools}
\newacronym{api}{API}{Application Programming Interface}
\newacronym{mib}{MIB}{Mission Information Base}
\newacronym{asi}{ASI}{Italian Space Agency}
\newacronym{ap}{AP}{Activity Plan}
\newacronym{tc}{TC}{Telecommand}
\newacronym{tm}{TM}{Telemetry}
\begin{document}
\let\WriteBookmarks\relax
\def\floatpagepagefraction{1}
\def\textpagefraction{.001}
\shorttitle{Operations and ground segment of the \gls{cheops} mission}
\shortauthors{A. Heitzmann et~al.}

%%% TITLE SUGGESTIONS %%%%%
% \title [mode = title]{CHEOPS Ground Segment. Overview of Mission \& Science Operations subsystems}   
\title [mode = title]{CHEOPS Ground Segment: Systems and Automation for Mission and Science Operations}

% \tnotetext[1]{This document is the results of the research
%    project funded by the National Science Foundation.}

% \tnotetext[2]{The second title footnote which is a longer text matter
%    to fill through the whole text width and overflow into
%    another line in the footnotes area of the first page.}

% \author[1,3]{J.K. Krishnan}[type=editor,
%                         auid=000,bioid=1,
%                         prefix=Sir,
%                         role=Researcher,
%                         orcid=0000-0001-0000-0000]
% \cormark[1]
% \ead{jkk@example.in}
% \ead[url]{www.jkkrishnan.in}

\affiliation[1]{organization={Department of Astronomy, University of Geneva},
                addressline={Chemin Pegasi 51}, 
                city={Versoix},
%               citysep={}, % Uncomment if no comma needed between city and postcode
                postcode={1290}, 
                country={Switzerland}}

\affiliation[2]{organization={INTA - Instituto Nacional de Técnica Aeroespacial, CEIT - Centro Espacial INTA Torrejón},
                addressline={Carretera de Ajalvir km. 4}, 
                city={Torrejón de Ardoz},
%               citysep={}, % Uncomment if no comma needed between city and postcode
                postcode={28864}, 
                country={Spain}}     
\affiliation[3]{organization={ESA - European Space Agency, European Space Operations Centre - ESOC},
                addressline={Robert-Bosch-Str. 5}, 
                city={Darmstadt},
%               citysep={}, % Uncomment if no comma needed between city and postcode
                postcode={64293}, 
                state={Hesse},
                country={Germany}}
\affiliation[4]{organization={Center for Space and Habitability, University of Bern},
                addressline={Gesellschaftsstrasse 6}, 
                city={Bern},
%               citysep={}, % Uncomment if no comma needed between city and postcode
                postcode={3012}, 
                country={Switzerland}} 
\affiliation[5]{organization={Weltraumforschung und Planetologie, Physikalisches Institut, University of Bern},
                addressline={Gesellschaftsstrasse 6}, 
                city={Bern},
%               citysep={}, % Uncomment if no comma needed between city and postcode
                postcode={3012}, 
                country={Switzerland}} 

\affiliation[6]{organization={Centre Vie dans l’Univers, Facult\'e des sciences, Universit\'e de Gen\`eve},
                addressline={Quai Ernest-Ansermet 30}, 
                city={Gen\`eve 4},
%               citysep={}, % Uncomment if no comma needed between city and postcode
                postcode={1211}, 
                country={Switzerland}}
\affiliation[7]{organization={ETH Zurich, Department of Physics},
                addressline={Wolfgang-Pauli-Strasse 2}, 
                city={Zurich},
%               citysep={}, % Uncomment if no comma needed between city and postcode
                postcode={CH-8093}, 
                country={Switzerland}}
\affiliation[8]{organization={ISDEFE - Ingeniería de Sistemas para la Defensa de España},
                addressline={Calle de Beatriz de Bobadilla 3}, 
                city={Madrid},
%               citysep={}, % Uncomment if no comma needed between city and postcode
                postcode={28040}, 
                country={Spain}}   
\affiliation[9]{organization={Procesia Proyectos y Servicios},
                addressline={Calle de Guzmán el Bueno 133}, 
                city={Madrid},
%               citysep={}, % Uncomment if no comma needed between city and postcode
                postcode={28003}, 
                country={Spain}}

\author[1]{Alexis Heitzmann\,$^{\href{https://orcid.org/0000-0002-8091-7526}{\protect\includegraphics[height=0.19cm]{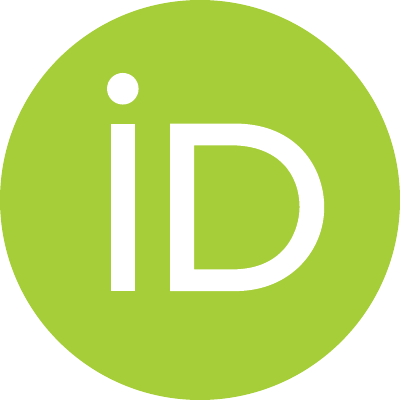}}}$}
\credit{Software, Writing - Original Draft, Writing - Review \& Editing, Visualization, Data curation.}
\cormark[1]
\fnmark[1]
\ead{alexis.heitzmann@unige.ch}

\author[9]{Anthony G. Maldonado}
\credit{Software, Writing - Original Draft, Writing - Review \& Editing, Visualization, Data curation}
\ead{cheops-moc@unige.ch}

\author[1]{Anja Bekkelien\,$^{\href{https://orcid.org/0000-0003-3778-3427}{\protect\includegraphics[height=0.19cm]{orcid.pdf}}}$}
\credit{Software, Writing - Original Draft, Writing - Review \& Editing, Visualization, Data curation}

\author[1]{Babatunde Akinsanmi\,$^{\href{https://orcid.org/0000-0001-6519-1598}{\protect\includegraphics[height=0.19cm]{orcid.pdf}}}$}
\credit{Writing - Review \& Editing}

\author[2]{Nuria {Alfaro Llorente}}
\credit{Project administration}

\author[7]{Mathias O. W. Beck\,$^{\href{https://orcid.org/0000-0003-3926-0275}{\protect\includegraphics[height=0.19cm]{orcid.pdf}}}$}
\credit{Project administration, Software, Writing - Review \& Editing}

\author[1]{Nicolas Billot\,$^{\href{https://orcid.org/0000-0003-3429-3836}{\protect\includegraphics[height=0.19cm]{orcid.pdf}}}$}
\credit{Project administration, Software, Writing - Review \& Editing}

\author[4,5]{Christopher Broeg\,$^{\href{https://orcid.org/0000-0001-5132-2614}{\protect\includegraphics[height=0.19cm]{orcid.pdf}}}$}
\credit{Project administration, Writing - Review \& Editing}

\author[1]{Adrien Deline}
\credit{Project administration, Software, Writing - Review \& Editing}
\ead{cheops-pso-members@listes.unige.ch}

\author[1,6]{David Ehrenreich\,$^{\href{https://orcid.org/0000-0001-9704-5405}{\protect\includegraphics[height=0.19cm]{orcid.pdf}}}$}
\credit{Project administration}

\author[2]{Naiara {Fernández de Bobadilla Vallano}}
\credit{Software, Writing - Original Draft, Writing - Review \& Editing, Visualization, Data curation}

\author[4,5]{Andrea Fortier\,$^{\href{https://orcid.org/0000-0001-8450-3374}{\protect\includegraphics[height=0.19cm]{orcid.pdf}}}$}
\credit{Project administration}

\author[2]{María {Fuentes Tabas}}
\credit{Software, Writing - Original Draft, Writing - Review \& Editing, Visualization, Data curation}

\author[2]{María J. {González Bonilla}}
\credit{Project administration, Writing - Review \& Editing, Visualization}

\author[3]{Marcus G. F. Kirsch}
\credit{Project administration, Writing - Review \& Editing}

\author[1]{Monika Lendl\,$^{\href{https://orcid.org/0000-0001-9699-1459}{\protect\includegraphics[height=0.19cm]{orcid.pdf}}}$}
\credit{Project administration, Writing - Review \& Editing}

\author[8]{David {Modrego Contreras}}
\credit{Project administration, Software, Writing - Original Draft, Writing - Review \& Editing, Visualization, Data curation}

\author[2]{Eva M. {Vega Carrasco}}
\credit{Project administration}


\cortext[cor1]{Corresponding author}
% \cortext[cor2]{Principal corresponding author}

\begin{abstract}
The \gls{cheops}, the first \gls{esa} small-class mission, has been performing photometric astronomical observations with a particular emphasis on exoplanetary science for the past five years. A distinctive feature of \gls{cheops} is that the responsibility for all operational aspects of the mission lies with the consortium rather than \gls{esa}. As a result, all subsystems, their architecture, and operational processes have been independently developed and tailored specifically to \gls{cheops}. This paper offers an overview of the \gls{cheops} operational subsystems, the design, and the automation framework that compose the two main components of the \gls{cheops} ground segment: the \gls{moc} and the \gls{soc}. This comprehensive description of the \gls{cheops} workflow aims to serve as a reference and potential source of inspiration for future small and/or independent space missions.

\vspace{1cm}\LARGE\color{red}

\end{abstract}

% \begin{highlights}
% \item Research highlights item 1
% \item Research highlights item 2
% \item Research highlights item 3
% \end{highlights}

\maketitle

\section{Introduction}

The CHaracterising ExOPlanet Satellite (\gls{cheops},\break
\citealt{Benz2021CHEOPS}) is the first small-class (S-class) mission of the \gls{esa} Cosmic Vision 2015–2025 programme. \gls{cheops} was born from a partnership between \gls{esa} and the \gls{cheops} Consortium led by the University of Bern, Switzerland. The mission's main scientific goal is the characterisation of exoplanets, via the method of exoplanet transits \citep{Winn2010}. Mission requirements were driven by the scientific needs to reach a photometric precision of 20~ppm (parts per million) in 6~hours of integration time for bright stars (corresponding to the transit of an Earth-size planet around a G5 dwarf star brighter than \nth{9} magnitude in the V~band), and 85~ppm in 3~hours integrations for stars of \nth{12} magnitude in the V~band (corresponding to a Neptune-size planet transiting a K-type star). \cite{Fortier2024perf} provide an in-depth analysis of \gls{cheops} performance throughout the first 3.5~years of operations~(2019-2023).

\gls{cheops} was launched on December \nth{18}, 2019 from Kourou, French Guyana on Soyuz flight VS23. The satellite platform is an AS-250 from Airbus Defence \& Space and the payload, developed under the responsibility of the University of Bern, is a Ritchey-Chr\'etien telescope with a primary mirror of effective diameter of 30~cm. Incoming light is fed to its single instrument, a high-performance photometer with a frame-transfer back-side illuminated charge-coupled device (CCD) in its focal plane. \gls{cheops} is nadir-locked in a Sun-synchronous Low-Earth orbit at an altitude of 690~km, and a local time of the ascending node of 6~a.m. This means that the orbital period of \gls{cheops} is 98.7~min, closely following the day/night terminator, maximizing the Sun exposition of the solar panels and minimizing Earth stray-light contamination. As a result of the orbit design and the \gls{gs} used to communicate with the satellite, the spacecraft is visible by the \gls{gs} approximately 4–6 times a day, 2–3 times per morning and evening. 

\gls{cheops}' ground segment is classically divided into two main entities. The \acrfull{moc}, located in the \gls{ceit}, at the \gls{inta}, Spain, handles telecommunication with the spacecraft, the satellite status monitoring, the orbit determination and control (\textit{i.e.}, collision avoidance, orbital correction manoeuvres, and de-orbiting at the end of the mission), and the attitude monitoring. The \acrfull{soc}, hosted by the astronomy department of the Université de Genève, Switzerland, is the bridge between the Project Science Office and the \gls{moc}. The \gls{soc} is responsible for handling observer requests, planning the sequence of scientific activities to be executed on board, as well as the data processing, distribution and archiving. Unlike other ESA missions, all aspects of mission operations for the \gls{cheops} mission fall under the responsibility of the Consortium. 
For this reason, \gls{cheops}' operational subsystems have been developed either from scratch following \gls{ecss} standards, or through the reuse, customization, or extension of existing systems. \gls{cheops}' S-class mission status imposed a total contribution (including launch), of 50 million\,\euro\ from \gls{esa} as well as a 4-year time frame for the development phase. The development was led by the University of Bern, with important contributions from members of the \gls{cheops} Consortium located in Austria, Belgium, France, Germany, Hungary, Italy, Portugal, Spain, Sweden, and the United Kingdom. The consortium contribution matched the \gls{esa} contribution, for a total of 105 million \euro\ covering design, fabrication, launch and operations for the nominal mission (which ended in 2023).

\begin{figure}[h! tbp]
	\centering
	\includegraphics[width=\columnwidth]{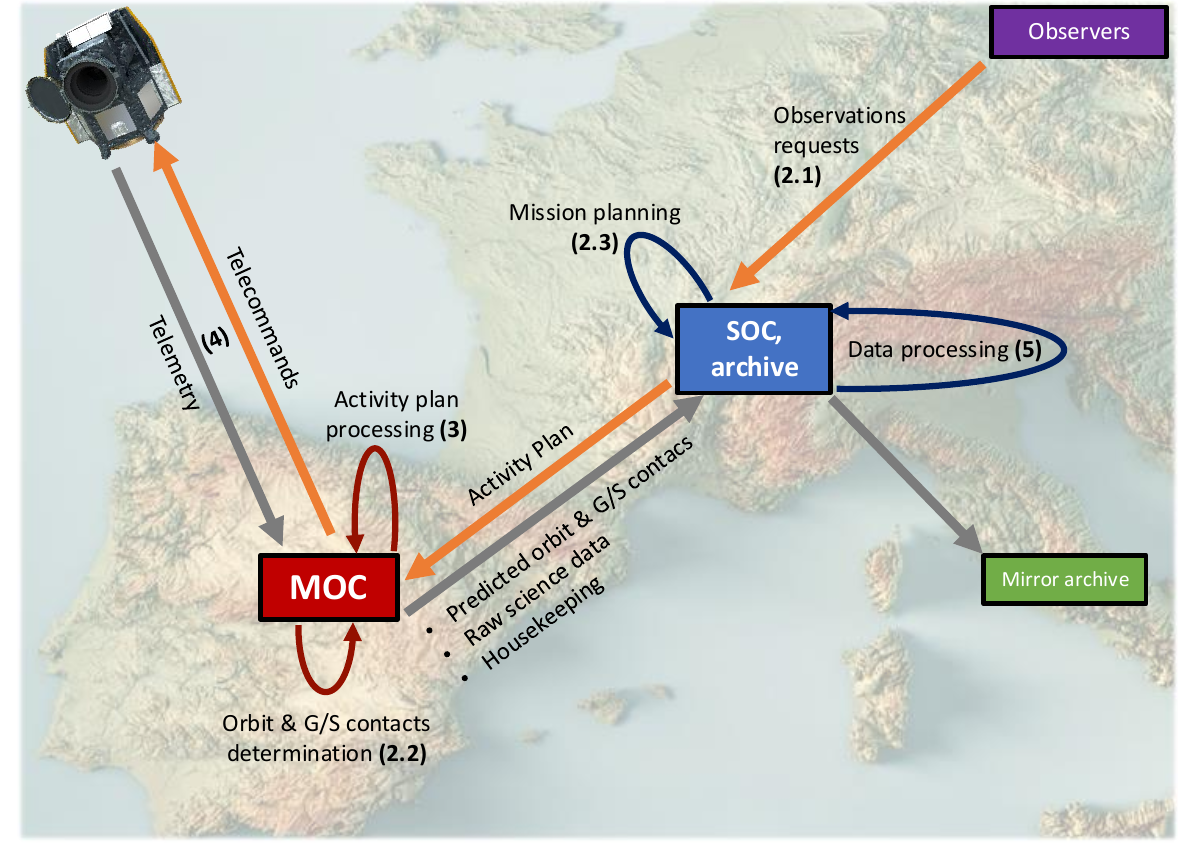}
	\caption{Overview of the \gls{cheops} operational chain. The ground segment components, \gls{moc} and \gls{soc}, are represented by coloured boxes. The orange arrows describe the weekly operations from reviewing and ingesting the observations requests to the uplink of the \gls{tc}s. The grey arrows trace the Science and Housekeeping data flow, from raw, to stored processed data accessible to the community, as well as the orbit and pass time determination, crucial for the planning phase. Finally, the steps outlined in this paper are marked with a bold number corresponding to their respective section.}
	\label{fig:overview_operations}
\end{figure}

The aim of this paper is to showcase the \gls{cheops} ground segment software, highlighting the effort to automate operations for a small-class mission designed to run on a modest budget. The uniqueness of the in-house development of most subsystems can serve as a benchmark for future mission development and operations. Figure\,\ref{fig:overview_operations} provides a high-level overview of the operational sites and information flow, to guide the reader through all components of the mission operations. 

Section\,\ref{sec:mission_planning} focuses on mission planning, where observing time requests from the scientific community, along with orbital and \gls{gs} contacts information provided by the \gls{moc}, are converted by the \gls{soc} into a sequence of activities, known as the \gls{ap}. Section\,\ref{sec:AP Process} then details how the \gls{moc} transforms the \gls{ap} into \acrlong{tc}s (\acrshort{tc}s) that will be executed onboard \gls{cheops}. Section\,\ref{sec:up_down_link} covers the operations at \gls{moc} during \gls{gs} contacts, when the \gls{tc}s are uplinked to the satellite and both the housekeeping and science data are downlinked, ingested in the \gls{moc} (housekeeping data only), and forwarded to \gls{soc}. This raw data is then processed, its quality assessed, and ultimately sent to the archive and the mirror archive, as outlined in Section\,\ref{sec:data_processing}. Finally, Section\,\ref{sec:additional_tools} presents some additional tools and software developed and used to complement the main \gls{cheops} operations.

\section{Mission Planning}
\label{sec:mission_planning}

Requests for the use of \gls{cheops} to observe targets of interest are made by members of the Science Team, Guest Observers, or the Instrument Team. In this section, the conversion of these requests to a sequence of instructions that are delivered to the spacecraft is described. First, the scientists submit \gls{or} to the \gls{soc} using the \gls{pht2}, described in Section\,\ref{subsec:OBStosoc}. \gls{soc} is also provided, by \gls{moc}, with the orbit files and the planned time of the passes (\gls{gs} contacts). The subsystems involved with these steps are the subject of Section\,\ref{subsec:MOCtoSOC}. The pool of existing \gls{or}s are then used to create an optimized observing plan, which is converted into the \acrlong{ap} (Section\,\ref{subsec:soctoMOC}), that \gls{soc} then transfers to \gls{moc}.

\subsection{Observing Requests from Users}
\label{subsec:OBStosoc}

The interface between the users (scientists) and the mission operations team is the \acrfull{pht2}\,\footnote{\href{https://cheops.unige.ch/pht2/}{https://cheops.unige.ch/pht2/}}, a web interface accessible to users who have been allocated observing time on \gls{cheops}. The software system is built with a MariaDB\footnote{\href{https://mariadb.com/}{https://mariadb.com/}} database server as the back-end, an Apache Web server running on CentOS with OpenSSL\footnote{\href{https://openssl-library.org/}{https://openssl-library.org/}} and PHP\footnote{\href{https://www.php.net/}{https://www.php.net/}}, and a front-end developed using WordPress\footnote{\href{https://wordpress.org/}{https://wordpress.org/}}. A screenshot of the web interface is shown on the bottom of Figure\,\ref{fig:mosaic_soc}. The PHP application connects to the MariaDB database through the MySQL Native Driver (mysqlnd, developed and maintained by the PHP project), which acts as the database client library.

New observation requests can be submitted using the User Interface or by submitting an XML file, containing all relevant information about an observation (details are outside the scope of this paper, but the interested reader is referred to the \gls{pht2} guidelines\footnote{\href{https://www.cosmos.esa.int/documents/1416855/16012138/PHT2-guidelines_AO-5_v_2.pdf}{PHT2 guidelines}}). All relative information for the users, targets, programmes and \gls{or}s are stored in the \gls{pht2} database. The \gls{pht2} also handles all status modifications from users or following observations (feedback from the \gls{mps}, subject of Section\,\ref{subsec:soctoMOC}). Every Monday, as part of the first step of the weekly planning phase, the \gls{soc} reviews all new, modified or stopped \gls{or}s. This review process validates all new \gls{or}s to make sure these are technically sound (\textit{e.g.}, target location or instrument settings) and that the observing strategy is properly implemented. XML files are generated for each new/modified \gls{or} and serve as input for the \gls{mps}.

\subsection{Predicted orbit and \gls{gs} contacts from \gls{moc}}
\label{subsec:MOCtoSOC}

The \gls{soc}'s \gls{mps} also requires the following information from \gls{moc} for the scientific planning generation: \gls{cheops} predicted orbit and planned \gls{gs} contacts. The creation of these files mainly involves the \gls{fds}, along with its automation, and the \gls{gs} Planner \gls{sw}.

The Flight Dynamics System (\gls{fds}) tool located at \gls{moc} is responsible for computing the satellite's orbit, derived products, orbital manoeuvres, attitude and pointing. The \gls{cheops} \gls{fds} is built on the Focus\footnote{\href{https://www.gmv.com/es-es/productos/espacio/focussuite}{https://www.gmv.com/es-es/productos/espacio/focussuite}} software suite, a collection of flight dynamics modules developed by GMV\footnote{\href{https://gmv.com/}{https://gmv.com/}}. The software is written in Fortran with a Java-based \gls{hmi}. Each module reads input files in ASCII format that can be read and edited in the \gls{hmi}. The input files contain configuration values and specify the paths for both internal and external files, including measurements, orbit, attitude, and event data. Focus generates the derived products needed for mission planning: the \gls{gs} contacts and the predicted orbit file. The generation of these products is based on the operational orbit, which contains the orbit from the entire mission. The operational orbit is updated twice a day after each orbit determination which is performed at the end of each session of passes, in the morning and evening. The key \gls{sw} modules involved in the orbit determination are detailed below:

\textbf{PREPRO} – It preprocesses tracking data files received from ground stations, which contain Doppler measurements recorded during \gls{cheops} passes. 
PREPRO converts two-way Doppler measurements (frequency shifts between the signal transmitted to and received from \gls{cheops}, in Hz) into radial velocity (km/s) and stores the observations in the NAPEOS (ESA's navigation tool for Earth orbiting satellites) Tracking Data Format to be used for orbit determination in the BAHN module.

\textbf{BAHN} – BAHN performs orbit determination using the Doppler measurements pre-processed with PREPRO. BAHN provides an estimation of the satellite's position and velocity in the J2000 reference frame at a user-defined reference epoch via a batch least-squares filter. If convergence is achieved, BAHN generates an orbit file including the determined period and a propagation six weeks in the future. This orbit is used to reconstruct the operational orbit.

Once the operational orbit has been updated with the latest orbit determination, it serves as input for the following \gls{sw} modules intended to generate the derived products: 

\textbf{EVENTS} – This component generates the \gls{gs} events with the visibilities between \gls{cheops} and the \gls{gs} available for the mission. The resulting event file includes potential passes for the subsequent 5 weeks.

\textbf{PROPAG} – Propagation of the orbit is done by means of a numerical integration method that calculates successive state vectors over a user-specified propagation interval. PROPAG makes use of solar and geomagnetic activity, Earth orientation parameters and perturbation models (geopotential, atmospheric, tropospheric and ionospheric effects). From an initial state vector included in the operational orbit, it propagates the orbit 4 months in the future.

The last step to generate the final \gls{fds} products consists of launching the PRODGEN module in Focus:

\textbf{PRODGEN} – From the outputs generated by the \gls{fds} modules previously described, PRODGEN generates the final products required by other subsystems, including:

\begin{itemize}
    \item iOrbitEvents: derived from the EVENTS file, this product contains visibility information needed at the \gls{ceit} \gls{gs} Planner for contact scheduling at the \gls{inta} \gls{gs} in coordination with other supported missions.
    
    \item Predicted Orbit: generated from the PROPAG orbit, it is sent to \gls{soc} and serves as input for the \gls{mps} for the \gls{ap} creation and occasionnal long-term planning.
\end{itemize}

The various Focus modules within \gls{fds} can be executed either manually, in isolation by users, or automatically in sequences. In the automated mode, the outputs of one module serve as inputs for subsequent modules, and the configuration values must be coordinated. This functionality is handled by an additional software component, Autofocus. Autofocus enables the automated execution of \gls{fds} workflows that involve the sequential operation of multiple modules. The workflows are encoded in the Spacecraft Operations Language (native to Autofocus), which facilitates configuration of the modules, initialization of their execution and retrieval of output information. This capability enhances operational efficiency and ensures consistency in the execution of complex flight dynamics processes. \newline

\label{subsec:CEIT G/S Planner}
For \gls{cheops} pass scheduling, the \gls{ceit} \gls{gs} Planner \gls{sw} tool, written in C\# and fully developed by \gls{inta}, is used. It understands the pass requests in different formats, graphically shows the current \gls{gs} schedule and the requests, detects and shows any conflict, modifies both the current schedule and the requests and saves and propagates any solution. The tool not only receives pass requests from \gls{cheops}, but also from other missions and generates a conflict-free solution. It also configures the control system of each \gls{gs} with the passes of all the missions including, the start and end times of the support, as well as configurations or the times of mode changes.
The Planner receives the iOrbitEvents file generated by the \gls{fds} and produces the GSAvailability, an XML file containing the planned \gls{gs} passes for \gls{cheops} for the following three weeks, specifically including the assigned \gls{gs}, start time and duration. \newline
The GSAvailability file is sent to \gls{soc} and serves as an input so the \gls{mps} can include pass related activities in the planning. This file is also processed internally at \gls{moc} as described in Section \ref{subsec:Prepass}.

\begin{figure}[h! tbp]
	\centering
	\includegraphics[width=\columnwidth]{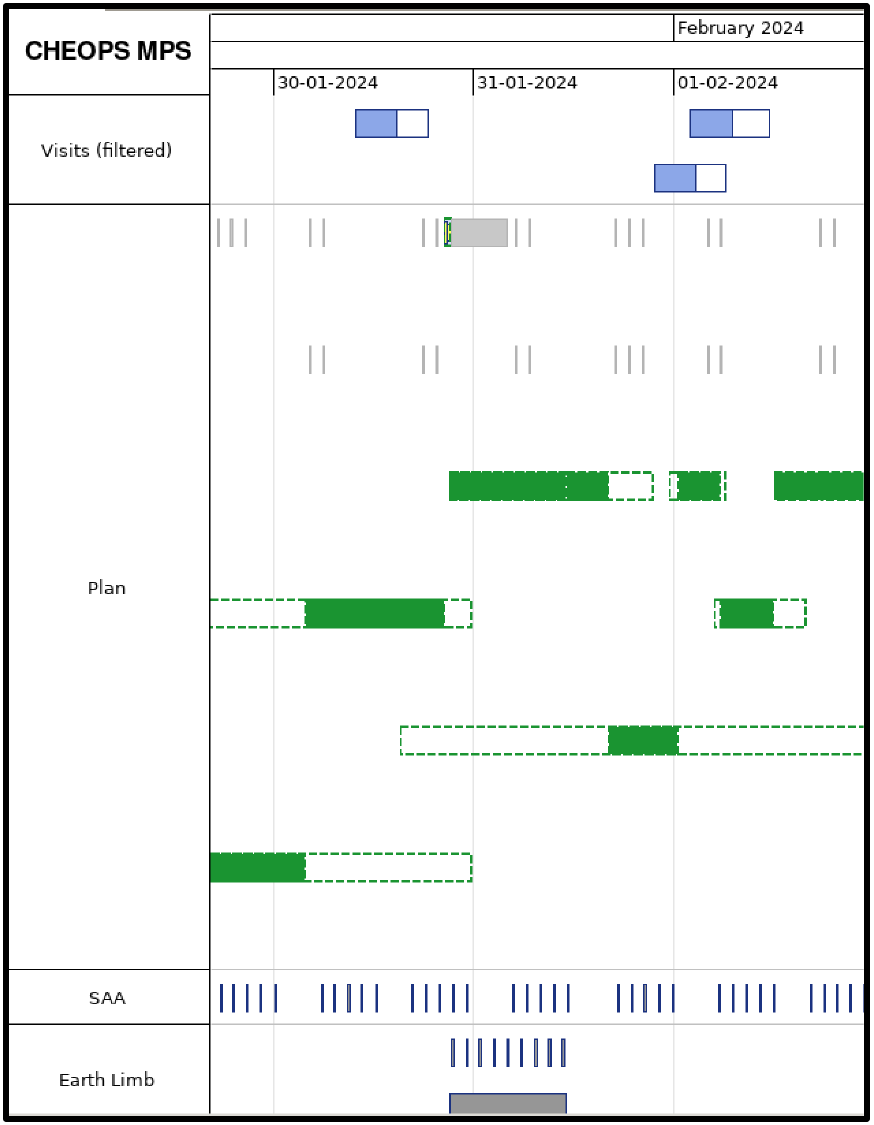}
	\caption{Screenshot of the \gls{mps} software developed by Deimos Engenharia, S.A. and used by \gls{soc} for the mission planning. Possible observations (called visits) are generated (top), then optimized into a plan by the Schedule solver (middle) taking into account scientific \gls{or}s and the constraints relative to \gls{cheops} orbit (bottom).}
	\label{fig:mps}
\end{figure}

\subsection{\acrlong{ap} creation at \gls{soc}}
\label{subsec:soctoMOC}

The main step of mission planning is to create an observation plan for \gls{cheops} and translating it into a sequence of procedures, the \acrfull{ap}, to be executed by the spacecraft. This process is carried out weekly at the \gls{soc} using the \acrfull{mps}, the software responsible for generating the observation schedule for the upcoming cycle, which typically spans one week beginning on Saturday.

The software was specifically created for \gls{cheops} and is maintained by Deimos Engenharia, S.A.\footnote{\href{https://deimos-space.com/}{https://deimos-space.com/}}.
The \gls{mps} reads all input (\textit{e.g.}, predicted orbit files, planned \gls{gs} contacts, configuration files, etc.) and output (\gls{ap}) files from an intray and an outray folders. The exchange of files between \gls{moc} and \gls{soc} is made via an FTP server and all files transferred through this FTP server are managed by the file dispatcher (described in Section \ref{sec:file_dispatcher}). The \gls{soc} operator creates the planning through the User Interface of the \gls{mps}, shown in Figure\,\ref{fig:mps}.

The source code of the \gls{mps} is written in Java and follows a client-server architecture. The client is based on a Deimos Engenharia, S.A. extended version of the Jaret Timebars library\footnote{\href{https://jaret.de/timebars/}{https://jaret.de/timebars/}}, combined with {Eclipse Rich Client Platform\footnote{\href{https://www.eclipse.org/}{https://www.eclipse.org/}}, and comprises a science functions submodule, that includes, \textit{e.g.} \gls{cheops} orbit propagations, that uses \gls{esa}'s Earth Observation CFI software\footnote{\href{https://eop-cfi.esa.int/}{https://eop-cfi.esa.int/}}. Communications between the client and the server use RESTful Web Services. The server has three submodules; the web services for the \gls{api} endpoints\footnote{Build with \href{https://cxf.apache.org/}{Apache CXF}}, a persistence module handling communication with the \gls{mps} database\footnote{Build with \href{https://hibernate.org/}{Hibernate ORM}}, and the Input/Output module\footnote{Build with the Java \href{https://github.com/nom-tam-fits}{Nom-tam-fits} library}, handling both XML and FITS\,\citep{FITS} files. All the necessary data required by the \gls{mps} are stored in the \gls{mps} database, which uses PostgreSQL\footnote{\href{https://www.postgresql.org/}{https://www.postgresql.org/}}.

As stated previously, the main role of the \gls{mps} is to generate the \gls{ap}. The process starts with the \gls{mps} generating all possible observations in a given time frame, accounting for all orbit related constraints. The goal is then to create an optimal sequence of observations, maximizing the time-critical \gls{ap} fill factor. This means optimizing a merit function, composed of: \acrfull{or} completion rate (currently unused), the ratio of Guaranteed Time to Guest Observers number of observations, the scientific priority of \gls{or}s, and the filling factor (minimizing \gls{cheops}' idle time). The optimization is performed by a genetic algorithm called the \textbf{Schedule Solver}. This piece of software was originally developed by Deimos Engenharia, S.A. and later improved by the \gls{soc} and is written in C. It is run \textit{outside} of the \gls{mps} software and takes as input a text file containing all the possible visits generated by the \gls{mps}. The Schedule Solver starts by creating multiple sequences of randomly chosen visits (\textit{i.e.} observations). During each iteration, the algorithm selects, combines, and mutates (applies random changes to) these visits sequences. These alterations are governed by parameters such as crossover and mutation probability. The cycle repeats for a predefined number of iterations, ultimately selecting the \gls{ap} with the highest overall ranking. The output from the schedule solver is a TXT file listing the optimal solution (a suite of observations) that can be ingested back in the \gls{mps}.
After human vetting of the solution by \gls{soc} and a validation by the Project Science Office of the observation plan (and of any proposed changes), the \gls{mps} converts the sequence of observations to the \gls{ap}. The \gls{ap} is an XML file listing all activities for \gls{cheops} to execute. These include slews between targets, target acquisitions, inhibition of data taking when crossing the South Atlantic Anomaly (region of high density of cosmic ray hits degrading the data quality), etc. These activities also specify the configuration of the instrument. Before sending the \gls{ap} to the \gls{moc}, additional files are generated with custom Python utility scripts. Among these are binary files called `StarMaps', used by the instrument to locate the correct star when pointing to the target location, and developed by the University of Vienna. All these files are pushed to the file dispatcher (described in Section\,\ref{sec:file_dispatcher}) which takes care of sending the files to \gls{moc} through a FTP server.

\section{\acrlong{ap} processing}
\label{sec:AP Process}
Upon reception of the \gls{ap} and the StarMaps at \gls{moc}, they are processed and translated into \gls{tc}s for uplink to the satellite during a scheduled pass. This processing task involves the coordination of multiple subsystems and software tools within the \gls{moc}, including the File Transfer System, Automation, \gls{mcs} and \gls{fds}, each of them playing a specific role in the workflow. Figure\,\ref{fig:MOC SW} presents these systems and the interface between them.

\begin{figure*}[h! tbp]
	\centering
	\includegraphics[width=\textwidth]{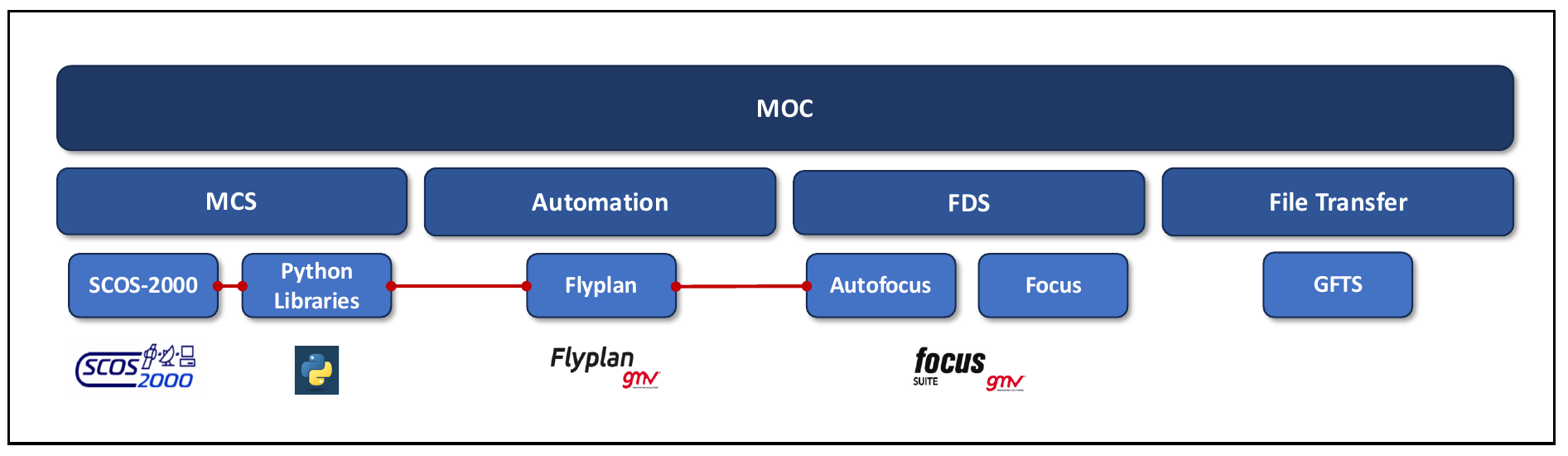}
	\caption{Overview of the \gls{moc} \gls{sw}. Files are handled by the \acrfull{gfts}, then detected by the automation system (orchestrated by Flyplan) which triggers the \acrfull{ap} processing. A sanity check of the \gls{ap} is performed by the \acrfull{fds} and if successful, the sequence of activities in the \gls{ap} are converted to \gls{tc}s by the \acrfull{mcs} before being uplinked to the spacecraft.}
	\label{fig:MOC SW}
\end{figure*}

\label{subsec:GFTS}
The initial step before the \gls{ap} processing is the reception of the files at \gls{moc}, a task handled by the \gls{gfts}. \gls{gfts} is a file transfer system used to automatically deliver files between hosts with or without a \gls{gfts} instance installed. This tool guarantees the orderly transfer of files among \gls{moc} components and external entities such as \gls{soc}, while also verifying the integrity of the received files. \gls{gfts} establishes the interface with \gls{soc} via FTP and initiates an action procedure to move the \gls{ap} and StarMaps files to the appropriate folder where the Automation system is able to detect and trigger the \gls{ap} processing tasks. 

In accordance with the S-class mission budget, the \gls{moc} is fully automated for routine tasks in order to minimize operator presence outside of working hours. All automated activities are orchestrated by the Flyplan software (provided by GMV) that allows automatic planning and execution of \gls{mcs} and \gls{fds} subsystems, offering a comprehensive Gantt-style visualization that provides insight into the schedule and reflects the execution status of all tasks (see Figure\,\ref{fig:AP_flyplan}). Its customizability allows the operator to create scheduling routines with Jython, the Python interpreter designed to run on the Java platform, for scripts in Python and Spacecraft Operations Language. A key component of Flyplan is the monitorCHEOPS activity: while some of the operations are time-driven, this script provides the automation system the capability to be data-driven whenever necessary, permitting flexible and independent operations. It monitors a configurable list of pathnames to detect input files and then triggers the corresponding processing task.

At this stage, the \textbf{monitorCHEOPS} component from Flyplan triggers the \gls{ap} processing in the different subsystems, as depicted in Figure\,\ref{fig:AP_flyplan}: first, an \gls{fds} sequence is launched in Autofocus to ensure the \gls{ap} complies with operational constraints, as detailed in Section\,\ref{subsec:CONCHECK}; second, a Python script in the \gls{mcs} is initialized to convert the \gls{ap} and the StarMaps into \gls{ap} \gls{tc}s calling the \textbf{CommandBuilder}; and, finally, one last Python script is triggered to assign the \gls{tc}s generated to the next available pass for its automatic uplink to the satellite. The last 2 steps are the focus of Section\,\ref{subsec:ACTPLA_conversion}.\newline

\begin{figure}[h! tbp]
	\centering
	\includegraphics[width=\columnwidth]{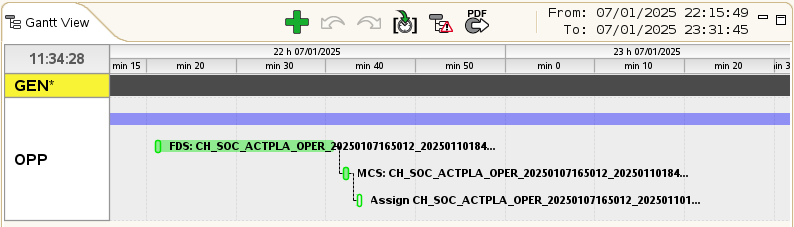}
	\caption{Flyplan Gantt chart with \gls{ap} processing activities in \gls{fds} and \gls{mcs} scheduled by the monitorCHEOPS upon reception of the \gls{ap} at \gls{moc}.}
	\label{fig:AP_flyplan}
\end{figure}

\subsection{Operational Constraints Check}
\label{subsec:CONCHECK}
The Autofocus sequence begins with the simulation of the satellite's attitude and the operational constraints compliance check. The following modules in the \gls{fds} support these tasks:

{\textbf{ATTSIM}} - It simulates the spacecraft attitude based on the activities included in the \gls{ap} and generates a predicted attitude file required for the subsequent module.

{\textbf{CONCHECK}} – Due to the physical and operational constraints placed on the satellite, the line of sight of the telescope must never be less than a defined Sun Exclusion Angle (initially set to 120 degrees and relaxed to 117 as of February 2025 to allow greater sky coverage) with respect to the Sun vector. This value is configurable by the user in the CONCHECK module, which is in charge of ingesting the \gls{ap} and checking the compliance of the constraint for every slew (transition between two \gls{cheops} pointings) detailed in the \gls{ap}. A constraint checker report indicating a Pass status is generated and disseminated via \gls{gfts} to the \gls{mcs} exclusively upon successful completion of this verification process.

\subsection{\gls{tc} Conversion and Pass Assignment}
\label{subsec:ACTPLA_conversion}

At this point, the \gls{ap} and the StarMaps are processed by the \acrfull{mcs}. \gls{cheops} \gls{mcs} is based on \gls{esa}'s SCOS-2000 (version 5.4.2.). SCOS-2000\,\citep{scos2000} is a spacecraft control system framework developed using an object-oriented analysis and design approach. Written in C++, it provides the basic processing expected of a spacecraft control system, which includes but is not limited to \gls{tm} processing and telecommanding. 

SCOS-2000 allows missions to configure the system for their requirements using system configuration options such as environment variables and the database component, the \gls{mib}. The latter involves defining mission-specific information, such as detailing every \gls{tc} and \gls{tm} packet.

Of particular relevance is the CommandBuilder, an \gls{mcs} tool that converts certain input files with activities into \gls{tc}s expressed in an \gls{mcs} legible format. The result is stored in an SSF file, an ASCII file that can be read by any standard SCOS-2000 using the \gls{mib}.

The \gls{mcs} automation capability is provided through a library of Python modules. This scripting layer grants access to all \gls{mcs} tasks, enabling interaction with SCOS-2000 and the execution of routine operations, such as processing the \gls{ap}. Specifically, when the time for a scheduled \gls{mcs} activity in Flyplan is reached, a Python script is triggered to verify the status of the constraint checker report and then invoke the CommandBuilder, where each activity is mapped to one or several \gls{tc}s using a configuration rule file. Moreover, the CommandBuilder estimates the \gls{ap} uplink duration and generates a Conversion Report file containing the conversion result which is then provided to \gls{soc} via \gls{gfts}.

The last activity consists of an additional Python script in charge of assigning the \gls{ap} SSF file to a pass for its uplink. Using the estimated uplink duration as input, this script selects the next pass in the current GSAvailability file with sufficient duration for the uplink and assigns the converted SSF file to it.

\section{\gls{cheops} passes: \gls{tc} uplink and \gls{tm} downlink }
\label{sec:up_down_link}
As previously mentioned, there are four to six contacts between \gls{cheops} and the \gls{gs} per day. Before the pass, the \gls{tc}s that will be automatically uplinked during the pass must be made available in the corresponding pass folder. Then, during the pass, the \gls{tc}s are uplinked to the satellite (including the \gls{tc}s from the \gls{ap}); the real-time \gls{tm} (housekeeping data) is received in the \gls{gs} and directly transferred to the \gls{mcs} so the operators can monitor the satellite high-level status if desired; and the recorded \gls{tm} is received in the \gls{gs} and recorded in a binary file. Finally, after the pass, the \gls{tm} file from the \gls{gs} is processed in the \gls{mcs} to generate separate \gls{tm} files, so the science and housekeeping data files can be transferred to \gls{soc}.

The \gls{tm} received from the satellite is converted into \gls{tm} packets and stored in the \textbf{\gls{parc}}. The \gls{parc} is used for the archiving of \gls{tm}, \gls{tc}, and event data. It stores the data in a MySQL database, allowing users to query the archive to retrieve data.

Several MOC subsystems and software tools are used for the pass related activities: the \gls{mcs} along with the Python scripts layer, the Automation, and the File Transfer System.

The following sections describe the main pass related activities (shown in Figure\,\ref{fig:Pass-related activities}) and the subsystems involved in more detail.

\begin{figure}[h]
    \centering
    \includegraphics[width=1\linewidth]{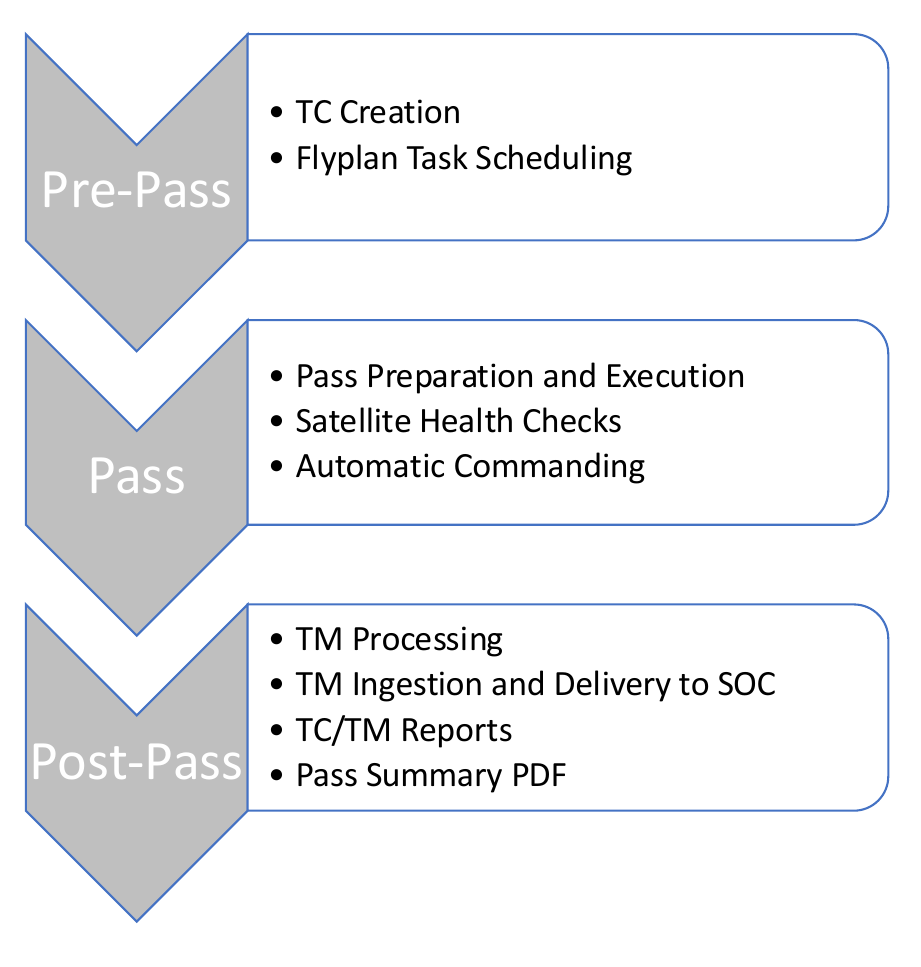}
    \caption{Pass-related activities overview.}
    \label{fig:Pass-related activities}
\end{figure}

\subsection{Pre-pass activities}
\label{subsec:Prepass}
Once the GSAvailability file is generated with the \gls{ceit} \gls{gs} Planner tool, it is detected by the monitorCHEOPS activity in Flyplan and processed. The GSAvailability file is processed through the SCOS-2000 CommandBuilder to create the \gls{tc}s that will be uplinked to the satellite. For every pass found in the file, \gls{tc}s stacks to begin and end the \gls{mcs}-\gls{gs} communication and to start and stop the downlink of the recorded \gls{tm} are generated. A pass folder with a specific identifier is also created for each pass, and the \gls{tc}s that have already been generated are assigned to the corresponding pass folder. 

Once all \gls{tc}s are created and assigned to their pass folder, the TaskInjector program is responsible for scheduling pass-related activities in Flyplan that are defined in a configuration file and are scheduled with respect to the pass start time. These activities include the pass activities (pass preparation and execution, satellite system health checks, and automatic commanding) and the pass processing activity.

\subsection{Pass activities}
\label{subsec:Pass}
Three main activities are performed during the pass. They are Python scripts orchestrated by Flyplan that are able to interact with SCOS-2000. They are part of the \gls{mcs} automation (Section\,\ref{subsec:ACTPLA_conversion}).

\subsubsection{\textbf{Pass preparation and execution}}
\label{subsubsec:Pass preparation and execution}
This \gls{cheops} activity starts a few minutes before the pass, and is responsible for establishing the \gls{mcs}-\gls{gs} connection, uplinking the \gls{tc}s assigned to the pass, and downlinking the \gls{tm}. 

First, the Python script monitors and restarts different SCOS-2000 tasks to ensure \gls{mcs} tools needed for the pass are up and running. Furthermore, it identifies the current pass folder, which contains the \gls{tc} stacks to be uplinked during the pass.

The \textbf{Auto Stack} is an \gls{mcs} application that allows the automation system to load and send pregenerated \gls{tc}s. For \gls{cheops}, two Auto Stack instances have been configured and reserved for the automated operations, enabling simultaneous tasks to be performed by the automation system.

At the beginning of the pass, the Python script loads in the Auto Stack and sends the \gls{tc}s found in the current pass folder to open the \gls{mcs}-\gls{gs} connection.

The \textbf{SMON} is an \gls{mcs} process that allows \gls{tm} parameter provision from the \gls{parc}. The Python script uses this process to retrieve the engineering value of a specific \gls{tm} from the \gls{parc} in different data formats, along with the time-stamp if desired. It is possible to retrieve the most recent value, any value in the past, or wait for the next sample to be received.

Thus, once the \gls{mcs} starts to receive real-time \gls{tm} from the satellite, the Python script waits for and retrieves the on-board time \gls{tm} value from the \gls{parc}. This \gls{tm} is used to calculate the accuracy of the Time Correlation coefficients. 

Modern spacecraft often carry a GPS receiver that provides an accurate synchronization between the spacecraft time and the ground time (usually synchronised to \gls{utc}) needed for time-tagging both \gls{tm}s and \gls{tc}s. Nevertheless, to reduce costs, \gls{cheops}' time is based on a local oscillator inside the On Board Computer. Therefore, an additional SCOS-2000 feature was configured for \gls{cheops} to accurately reconstitute the \gls{obt} following the \gls{ecss} time management system. This feature supports a linear approximation for a free-running clock, like the one used in \gls{cheops}, updating the transformation coefficients every time the monitored accuracy is lower than required.

The Python scripts also have the capability to get or modify SCOS-2000 MISC variable values. This feature allows the automation system to verify the ground commanding capability is available before sending the \gls{tc}s loaded in the Auto Stacks, to change the transmission mode to AD mode, and to check that the Auto Stack has completed sending the \gls{tc}s so the following can be loaded.

The \gls{tc}s loaded in the Auto Stacks are first sent to the \gls{gs} and then to \gls{cheops}. The AD mode is used for the communication, since it allows the detection of missing \gls{tc}s and their retransmission during the communication. Several steps of acknowledgement are enabled in order to trace both the reception and transmission of the \gls{tc}s by the \gls{gs}, and the reception and execution (if it is a \gls{tc} to be executed immediately) of the \gls{tc}s on board. Using this transmission  mode, SCOS-2000 verifies the correct and complete uplink of the \gls{tc}s.

Once the ground commanding capability is available, the remaining \gls{tc}s in the current pass folder are loaded one by one in the Auto Stacks and sent to \gls{cheops}. The \gls{ap} (in case it is assigned in the current pass folder, see Section \,\ref{subsec:ACTPLA_conversion}) and the \gls{tc}s to start and end the downlink of the recorded \gls{tm} are sent at this stage.

At the end of the pass, the \gls{tc}s to close the \gls{mcs}-\gls{gs} connection are loaded in the Auto Stack and sent.

\subsubsection{\textbf{Satellite systems health-checks} }

This activity at the beginning of the pass is intended to check the high-level status of different platform subsystems based on Flight Operational Procedures provided by the manufacturer. These Flight Operational Procedures are written in Python and invoked as external commands. These scripts wait for and read \gls{tm} values from SCOS-2000 (see Section \ref{subsubsec:Pass preparation and execution} for \gls{tm} parameter provision).

\subsubsection{\textbf{Automatic commanding}}

This activity is executed at the beginning of the pass, and offers two additional features that make use of the \textbf{SCOS-2000 External Interface}. The first is the real-time generation and transmission of \gls{tc}s to \gls{cheops} in response to specific \gls{tm} values, and the second is the detection of defined \gls{tm} packets received on ground so they can be retrieved from the \gls{parc} and decoded.

Among its uses is the recovery of \gls{tm} from previous passes in which gaps were detected. Given that only two \gls{tm} downlinks can be executed simultaneously, the automatic commanding Python script is responsible for checking that the previous \gls{tm} downlink has finished before commanding the next one. 

\subsection{\textbf{Post-pass activities}}
\label{subsec:Post Process}
The pass processing activity is responsible for processing the \gls{tm} recorded on board between consecutive \gls{gs} contacts that is downlinked during the pass. It enables the provision of science and housekeeping \gls{tm} to \gls{soc}, and the necessary \gls{tm} to monitor the satellite status to \gls{moc}. Furthermore, it allows \gls{moc} operators to verify the execution status of the different activities performed automatically during the pass.

This activity is orchestrated by Flyplan a few minutes after the pass, and consists of a series of Python scripts that perform the activities outlined below. These Python scripts are part of the \gls{mcs} automation (Section \,\ref{subsec:ACTPLA_conversion}).

First, the start time and the \gls{gs} data for both the current and previous pass are searched in the current GSAvailability file. This enables the generation of \gls{tm} reports covering the period from the beginning of the previous pass to the beginning of the current pass.

Secondly, the binary \gls{tm} file is fetched from the \gls{gs} via FTP connection so it can be processed in the \gls{mcs} and transferred to \gls{soc}:

\begin{itemize}
    \item The \textbf{SCOS-2000 \gls{tm} Processor} application is used to preprocess the \gls{gs} files in order to generate separate \gls{tm} files, one per Virtual Channel (real-time housekeeping data, recorded housekeeping data, and science data).

    \item The \gls{tm} files with housekeeping and science data are transferred to \gls{soc} via \gls{gfts}.

    \item The\textbf{ SCOS-2000 \gls{tm} Replayer} application is used to ingest the recorded housekeeping \gls{tm} into the \gls{parc}.

    \item The \gls{tm} files are checked to identify \gls{tm} gaps by detecting a jump in the Transfer Frames header Virtual Channel frame counter. If any gap is discovered, the gap information is included in a pending \gls{tm} gaps file.
\end{itemize}

Later, two additional products are generated and sent to \gls{soc} via \gls{gfts}:

\begin{itemize}
    \item The TCReport file to inform the \gls{mps} of the execution status of the \gls{tc}s associated to the activities programmed by the \gls{mps}. This file is generated using the \textbf{SCOS-2000 RGTgenerator} application, that processes the Conversion Report file generated by the CommandBuilder during the \gls{ap} processing, and retrieves the execution status of the \gls{tc}s from the \gls{parc}. The result is an XML file in Earth Explorer format readable by the \gls{mps}.

    \item The \gls{obt}-\gls{utc} report file, which is an XML file that includes the Time Correlation coefficients for the conversion of the \gls{tm} timestamp from \gls{utc} to \gls{obt} by \gls{soc}.
\end{itemize}

Additionally, some products are created to facilitate the monitoring of the satellite status by \gls{moc} operators, and to feed the \gls{fds} subsystem: 

\begin{itemize}
    \item Reports on \gls{tc} history, on-board events, and out-of-limits. They are created using SCOS-2000 executable programs that enable the printing of data shown in the \gls{mcs} displays over a specified time period in an ASCII file.

    \item \gls{tm} reports in ASCII or XML format. The \textbf{SCOS-2000 HKTMreporter} program, developed for \gls{cheops}, generates these reports using \gls{parc} \gls{tm} packets. This provides the \gls{fds} with the \gls{tm} required for some computations, for instance, the comparison of the real and simulated attitude in order to detect any deviations. Moreover, it enables the generation of trend plots for specific \gls{tm}s defined by the platform manufacturer.
\end{itemize}

Afterwards, different checks are performed in order to ensure that the \gls{gfts} is operating properly so that the files have reached their destination.

Finally, a PDF detailing the results of the different pass-related activities is generated using the Python library PyFPDF. The different Python scripts read the logs and reports generated during the pass and post-pass activities. A colour-coding system is used to display the result of the different activities, along with a few warning messages in case of failure. This PDF, as well as the most relevant logs and reports from the pass, are forwarded to the \gls{moc} operators using the Python smtplib and email libraries.
\newline

\section{Data processing at \gls{soc}}
\label{sec:data_processing}

When the housekeeping \gls{tm} and science data arrive at the \gls{soc}, the files are automatically detected by the system and ingested into the \gls{soc} data processing pipeline, which is shown in Figure\,\ref{fig:soc_pipeline} and described in more detail in the following sections. The pipeline converts all data to FITS format and reduces the science data to photometric light curves. It also produces data quality reports for the \gls{soc} operator per-downlink. The end data products are ingested into the \gls{cheops} Mission Archive (see Section\,\ref{subsec:archive}) that is hosted by the \gls{soc} and, for backup purposes and redundancy, replicated in a mirror archive hosted by the Space Science Data Center facility of the Italian Space Agency. The data products are distributed to the end users via the archive's web interface, the Archive Browser\footnote{\href{https://cheops.unige.ch/archive_browser}{https://cheops.unige.ch/archive\_browser}}.

Importantly, the entirety of the \gls{soc} processing pipeline (i.e., the content of Figure\,\ref{fig:soc_pipeline}) is, except for the ingestion of the quality reports in the \gls{mps}, fully automated and does not require any human intervention.

\subsection{Design decisions}

Due to the very restricted development time and cost imposed by the small-class mission type, the \gls{soc} was designed to be as simple and light-weight as possible. To achieve this, the \gls{soc} relied heavily on having collaborators with prior experience in designing and implementing \gls{soc} software for space missions.

To keep physical hardware to a minimum, the system was fully virtualized. This allowed it to run on existing department-level hardware infrastructure that is shared with other projects while benefitting from infrastructure maintenance and upgrades in a transparent manner. \gls{soc} would also reuse existing project management and software development tools maintained by the department, including a centralized Subversion version control system and a Jenkins server\footnote{\href{https://www.jenkins.io/}{https://www.jenkins.io/}} for automatic regression testing.

Due to the aforementioned restrictions, the priority was first placed on implementing the functionalities of the data processing and archival system while in a second phase focusing on streamlining the user interfaces and graphical representations of the system. The system would also reuse established and proven software libraries developed for other missions when this was possible. 

Lastly, the data processing and data archival system was designed to run in a fully automated and data-driven fashion in order to minimize the operational workload.

\subsection{Internal file transfer}
\label{sec:file_dispatcher}

The \gls{soc} system consists of multiple hosts, described in \ref{subsec:HWSOC}. The file transfer between the hosts is ensured by a Python script, the \textbf{file\_dispatcher}. The script uses a configuration file that defines the source and destination hosts and directories, the file types to be transferred to a given destination as well as optional pre- and post-transfer procedures that are generally used to back up transferred files or to delete unused files. The file\_dispatcher is executed at regular intervals by a cron job to continuously monitor directories for new files and trigger file transfers.

\subsection{Core libraries}

\textbf{common\_sw} provides interfaces, algorithms and other resources that are common to all software modules in the pipeline. Most notably, common\_sw provides the FITS data model which defines the structure of the data products that are produced by the pipeline and distributed to the end users\footnote{\href{https://www.cosmos.esa.int/web/cheops-guest-observers-programme/data-products-definition-document}{\gls{cheops} Data Products Definition Document}}. The definitions are in XML format and specify the structure of the header and data unit of all FITS extensions. C++ classes are generated from the XML definitions and contain accessor methods for modifying the content of the header and data unit, as well as functionality for reading and writing to disk, implemented using the \gls{nasa} CFITSIO library\,\citep{CFITSIO}. Furthermore, a Python interface to the C++ classes is generated using Boost Python\footnote{\href{https://www.boost.org/doc/libs/1_87_0/libs/python/doc/html/article.html}{Building Hybrid Systems with Boost.Python}}, an open source library for exposing C++ classes and functions to Python and vice versa. These classes ensure that all data products created by the pipeline adhere to the definitions across all subsystems. common\_sw also provides infrastructure for auto-generating documentation, both the data product documentation for the end users, which uses the XSLT language to transform the XML definitions to HTML, as well as code documentation for the developers using the documentation generator Doxygen\footnote{\href{https://www.doxygen.nl}{Doxygen}}.

\textbf{science\_tm} is a C++ library that provides low-level routines for compressing and decompressing the data of the science \gls{tm} received from \gls{moc} and is used by the Preprocessing chain when converting \gls{tm} to FITS images. 

\textbf{xml\_schema} is a collection of XML schemas that define file interfaces between \gls{moc} and \gls{soc}, as well as \gls{soc}-internal interfaces. The pipeline software uses CodeSynthesis XSD\footnote{\href{https://www.codesynthesis.com/products/xsd/}{CodeSynthesis XSD}}, an XML data binding compiler, to generate C++ bindings for the schemas that are used at run-time to read XML files. 

\begin{figure}[h! tbp]
	\centering
	\includegraphics[width=\columnwidth]{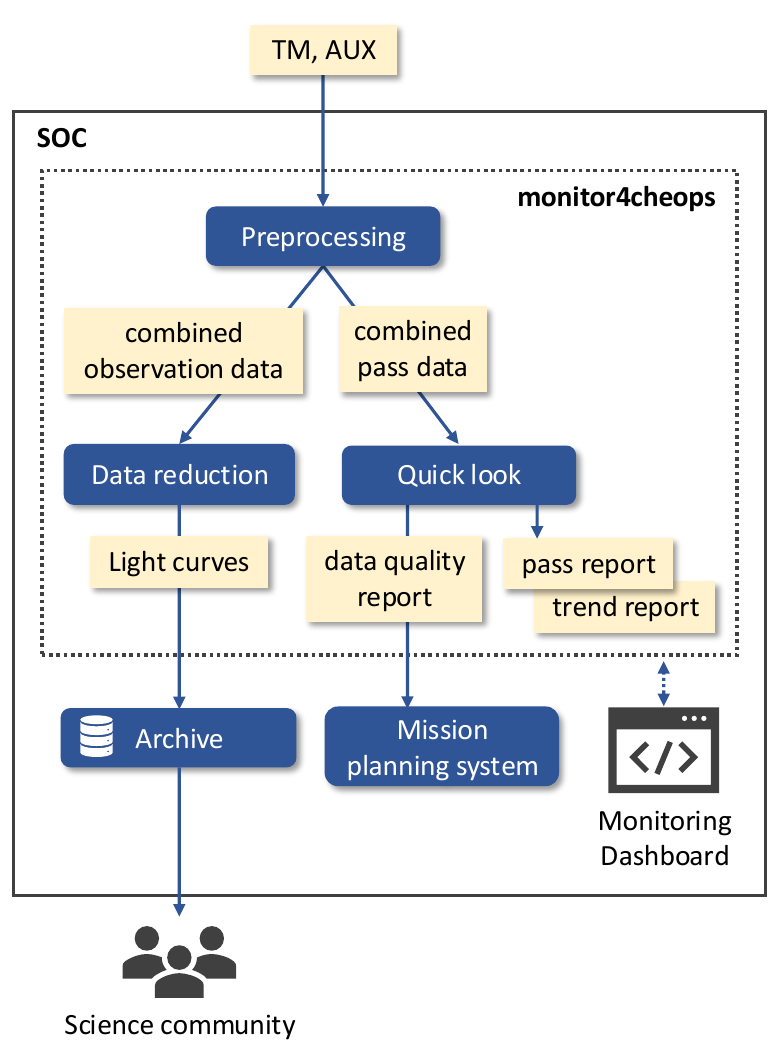}
	\caption{Logical overview of the subsystems (in blue) and data flow in the \gls{soc} data processing and archival system.}
	\label{fig:soc_pipeline}
\end{figure}

\subsection{Data processing pipeline}

Monitor4cheops is the component responsible for orchestrating the execution of the data processing pipeline, implemented in C++. It runs as a background service and automatically triggers the processing when data files from \gls{moc} are ingested into the system. Monitor4cheops has a generic plug-in architecture where the individual steps in the pipeline, which are implemented as stand-alone executables, are plugged into the system via configuration files. Thus, monitor4cheops is separate from the processing pipeline itself, making for a flexible and scalable system. The plug-ins adhere to a common interface; as input, they receive an XML file that defines the values of the input parameters and the location of the input files, and as part of their output, they are required to create a special \textit{trigger} file that is used by monitor4cheops to chain the processing steps. 

Monitor4cheops maintains an internal file repository of all ingested files, including the output files from the processing steps. File metadata is stored in RAM, and also duplicated to XML files on disk, and is used for selecting input files for the processing steps. The metadata includes the observation ID, pass ID, and file version. The processing steps themselves are defined in configuration files that specify their order in the pipeline, their input file types, and input parameters.  

The processing steps are logically grouped into the Preprocessing, Quick Look and Data Reduction processing chains, which are described in more detail in the following sections and shown in Figure\,\ref{fig:soc_pipeline}.

\subsubsection{Preprocessing chain}
\label{subsec:preprocessing}
Preprocessing is the first chain to be executed and it converts the raw \gls{tm}, in the form of Consultative Committee for Space Data Systems\footnote{\href{https://public.ccsds.org/Pubs/133x0b2e2.pdf}{CCSDS space packet protocol}} packets, as well as auxiliary data in XML format, to FITS data products. Preprocessing consists of two components: 

The \textbf{Conversion} component decodes and reformats the \gls{tm} according to the spacecraft's \gls{mib} and performs consistency checks to detect and report missing packets. The data is time-tagged using \gls{obt} to \gls{utc} correlation. Conversion calls the \textbf{science\_tm} library to decompress science images and finally converts the packets to FITS format. 

The \textbf{Combination} component is responsible for grouping FITS data by observation, which is done in two steps. First, by grouping the converted pass data and then, once all data of a given observation has been downlinked, which may take multiple passes (possibly days), by grouping the per-pass data. The former is input to the Quick Look chain, while the latter is input to the Data Reduction chain.

Preprocessing is implemented in C++ and uses the \gls{nasa} SPICE library\,\citep{SPICE} for attitude computation.

\subsubsection{Quick Look chain}
\label{subsec:quick_look}
The Quick Look chain, developed by the University of Cambridge, produces data quality reports that are used by the \gls{soc} operator and Instrument Team to monitor the status of observations and the instrument. It derives quality control parameters from images, image metadata, housekeeping, and auxiliary data and produces two report types: i) pass reports giving an immediate overview of the quality of the observational data in the pass, including technical quality, spacecraft pointing, source detection, and visualization of selected science images, and ii) trend reports generated at regular intervals for monitoring the long-term evolution of housekeeping parameters. The trend report generation is a two-step procedure. The housekeeping parameters received since the last report are first aggregated by binning and averaging the values and writing the aggregated points to disk. Then, the report is generated by selecting aggregated points for a given time interval. 

In addition, Quick Look produces technical quality parameters of observations that are fed back to the \gls{mps} to assess the efficiency of the observation.

Quick Look was designed to be flexible with regard to the selection of reported parameters to accommodate changes needed as a result of the evolution of instrument characteristics over the lifetime of the mission. The parameters configuration was therefore externalized to reference files.

\subsubsection{Data Reduction chain}

The Data Reduction is the last chain in the pipeline (see Figure\,\ref{fig:soc_pipeline}), responsible for creating the photometric light curves. The Data Reduction is extensively described in \cite{Hoyer2020drp}. It is implemented entirely in Python and consists of the following modules:

\textbf{Calibration}, which transforms the raw images output by Preprocessing into photo-electrons calibrated images.

\textbf{Correction}, which corrects the calibrated images for environmental effects including smearing, bad pixels, and background and stray light pollution.

\textbf{Photometry}, which performs aperture photometry on the corrected images and creates the final light curves.

\subsection{Monitoring Dashboard}
\label{subsec:monitoring_dashboard}
The Monitoring Dashboard is a web application that displays a graphical view of the processing pipeline and provides access to reports and log files created by the processing steps. It is the primary tool of the \gls{soc} operator for monitoring data processing and troubleshooting anomalies. The main view is an interactive timeline that lists the passes and observations and shows the status of the various processing chains. Clicking on a processing chain opens a window with detailed information about the execution of its processing steps. Both these interfaces can be seen on the right of Figure\,\ref{fig:mosaic_soc}.

The application is implemented in Python using the open source Django web framework\footnote{\href{https://www.djangoproject.com/}{https://www.djangoproject.com/}}. The information displayed in it is taken from an SQL database that is populated by monitor4cheops.

\begin{figure*}[h! tbp]
	\centering
	\includegraphics[width=\textwidth]{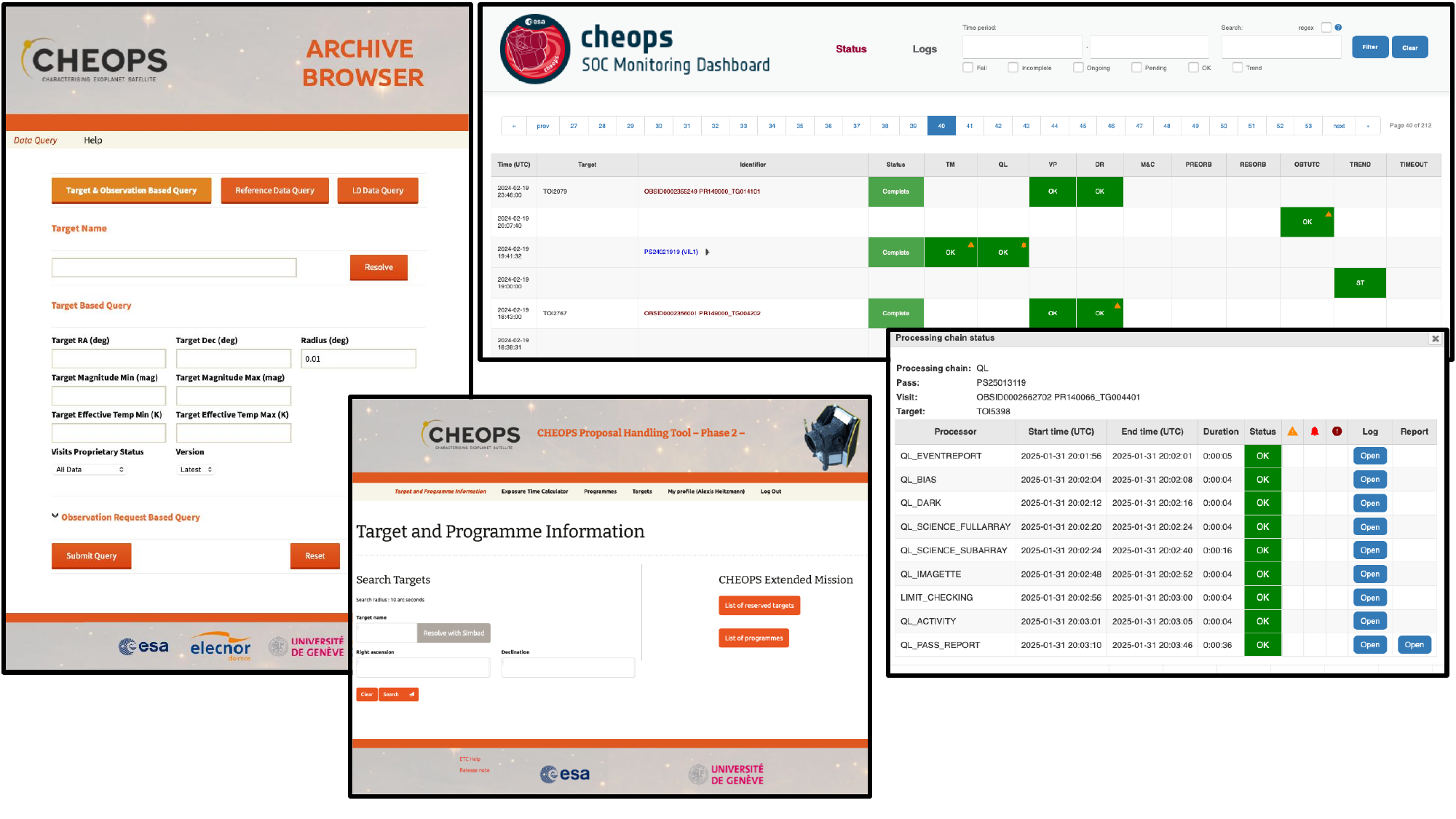}
	\caption{Mosaic of some of the \gls{soc} tools. \textbf{Top left}: Archive browser (Section\,\ref{subsec:archive}), allowing users to retrieve their data. \textbf{Top right}: Monitoring dashboard (Section\,\ref{subsec:monitoring_dashboard}), tracking the progress and logs of all the data processing pipeline. This is used daily by \gls{soc} to assess data quality, catch errors in data processing/reduction or identify missing data. The snippet on the right is a detailed view of all the `processors' in a specific module of the pipeline, here the Quick Look (Section\,\ref{subsec:quick_look}). \textbf{Bottom center}: Screenshot of the \gls{pht2} tool (Section\,\ref{subsec:OBStosoc}), where all the users \gls{or}s are entered and checked before being sent to the \acrlong{mps}.}
	\label{fig:mosaic_soc}
\end{figure*}

\subsection{Mission Archive}
\label{subsec:archive}
When an observation has been fully processed, the resulting data products are ingested into the Mission Archive and made available for download via the archive's web interface. The data owner is automatically notified by email of the availability of his data products.

The archive consists of two components: the archive server and the web interface, the Archive Browser, of which a screenshot is shown in Figure\,\ref{fig:mosaic_soc}. The archive server is based on a pre-existing generic data archival solution for space missions developed by Deimos Engenharia, S.A. A thin layer of customization was built on top of the generic solution to adapt it to the \gls{cheops} data model. The Archive Browser is a WordPress based web site that was developed specifically for \gls{cheops} as an interface between the archive server and the end users. The users can query for data using attributes related to the target and/or the observation. Proprietary data can only be downloaded by the data owner, but once the proprietary period has expired (generally one year after the observation), the data become available to all users.

Ten years after the end of the mission, all the data will be migrated to the \gls{esa} archive, guaranteering long term availability of all the \gls{cheops} products.

\subsection{Outlook}

The \gls{soc} system has been running on CentOS 7 since before the launch of \gls{cheops} in 2019. A global system upgrade was planned in 2024, the year when this CentOS version reached end-of-life. As Red Hat had at that point discontinued the development of CentOS in favour of CentOS Stream, AlmaLinux 9, which is binary-compatible with Red Hat Enterprise Linux, was selected as its replacement. The migration entailed upgrades of important third party dependencies, such as Python, C++, Boost and Postgresql, which in turn necessitated updates of the \gls{soc} software.

At the time of writing in early 2025, the migration of the compute node and the mission planning system server has been completed and the migration of the Mission Archive is ongoing. The archive software is tightly coupled with the OS and it was deemed too costly to migrate the software itself to AlmaLinux. Instead, the \gls{soc} has opted to containerize the archive (using podman\footnote{\href{https://podman.io/}{https://podman.io/}}) to allow the original software to run on the upgraded hosts.

\section{Additional tools}
\label{sec:additional_tools}

This section provides short descriptions of additional software tools used to assist in \gls{cheops} Operations.
\subsection{Parameter Archive and Plotting tool}
\gls{esa}'s \gls{must} and its web-based interface, Web\gls{must}, are frequently utilized for \gls{tm} visualization and trend analysis as alternatives to SCOS-2000, which is limited in certain functionalities. Originally implemented at the European Space Operations Center, the tool was later transferred to \gls{inta} and Universität Bern due to maintenance considerations. Currently, its operations and maintenance are managed collaboratively by the operations and instrument teams with the support of Solenix\footnote{\href{https://www.solenix.ch/products/must}{https://www.solenix.ch/products/must}}. The \gls{must} database is updated with RAPID files generated by SCOS-2000. For periodic trend analyses, the Python-based \gls{must}Link \gls{api} is employed. This \gls{api} facilitates authenticated server access and enables efficient data retrieval by utilizing a paginated approach, which divides large requests into smaller subsets to prevent server overload and ensure robust performance.

\subsection{CHEOPS Spacecraft Simulator}
The spacecraft simulator, based on \gls{esa}'s Simulus 5.4 framework and the SIMSAT platform, was specifically tailored for the \gls{cheops} mission. It provides an accurate representation of the spacecraft's on-board software. Although it lacks certain payload modelling capabilities, the simulator has been instrumental in validating both existing and new operational procedures. It played a critical role during pre-launch simulation campaigns and continues to support training activities and contingency operations.
The simulator features generic models that encompass \gls{tm} and \gls{tc} encoding and decoding, thermal network simulation, and communication with ground stations via the Space Link Extension protocol. Configuration loading and autonomous simulation of specific operations, such as \gls{gs} operator tasks, are facilitated through scripting in Java.

\subsection{CHEOPSim}

The \gls{cheops} simulator, described in \cite{Futyan2020cheopsim}, is a stand-alone \gls{soc} subsystem that creates accurate simulations of data from the payload, in the same format as the output from the Preprocessing chain (see Section\,\ref{subsec:preprocessing}). This tool was primarily used prior to routine operations and its main purposes were to create input data for testing and validating the \gls{soc} processing chains and to aid in the preparation of observations. Users interact with CHEOPSim through a web interface that launches simulations on a computing cluster.

Although beyond the scope of this paper, it is of note that CHEOPSim was vital for the assessment of CHEOPS scientific and instrumental performance.

\subsection{CHEOPS Visibility Tool}

Because of the combination of \gls{cheops} Low Earth Orbit and specific pointing restrictions (Sun and Moon Exclusion Angles), observations are subject to interruptions, and specific line of sight are only visible seasonally. To help users know when their target is visible with \gls{cheops}, the \gls{soc} developed, in 2024, the \gls{cheops} Visibility Tool\footnote{\href{https://gitlab.unige.ch/cheops/CHEOPS_visibility_tool}{https://gitlab.unige.ch/cheops/CHEOPS\_visibility\_tool}}. This Python code takes as input visibility and efficiency (fraction of a visit that actually contains images, result of interruptions) tables pre-calculated using the \gls{mps} at different time of the year and for different line of sights. These tables are interpolated in space and time to provide the user with a sky map showing the cone of visibility of \gls{cheops} for any given day of the year, as well as the window of observability and the efficiency for specific targets of interest.

\section{Conclusions}

\gls{cheops} is the first \gls{esa} S-class mission,  dedicated to characterising exoplanets orbiting around nearby bright stars with exquisite precision, paving the way for flagship \gls{esa} and \gls{nasa} missions, such as PLATO\,\citep{PLATO} or the James Webb Space Telescope\,\citep{JWST}. 

\gls{cheops} ground segment is divided into two main entities: the \acrlong{moc} (near Madrid, Spain) and the \acrlong{soc} (near Geneva, Switzerland). This paper gives a comprehensive overview of all the subsystems, software composing the \gls{cheops} operations and how they relate to each other to provide smooth missions operations, translating scientist requests for observations using \gls{cheops} into a plan of activities to be executed in the satellite, and to downlink, process, and archive the data to make it available to the scientists. 

By design, an S-class mission must operate within a tight budget. From the very beginning, this constraint has driven the \gls{cheops} development and operations teams to prioritize automation of the ground segment processes to meet cost requirements. The automation framework, detailed throughout this paper, has been paramount to allow efficient operations. With the high level of control over the various ground segment subsystems provided by this automation, the routine operations can be performed smoothly without the need for staff members to be present outside working hours. This not only reduces workforce demands (3-4 staff members for \gls{soc} and 5 for \gls{moc} in the current setting) but also ensures consistency in the execution of complex processes, and lowers the likelihood of human error. As a result, the teams can focus on performing specialized non-routine manual tasks that require in-depth and specialised knowledge of the different subsystems, enhancing the automation system, and developing new features driven by the evolving needs of the scientific community, such as the relaxation of the Sun Exclusion Angle, or data taking inside the South Atlantic Anomaly

After the first 4 years of operations, the mission was granted a 3-year extension, and after five years of successful operations, it has become apparent that automation enables the fulfilment of the mission's operational requirements without compromising any of its aspects. \gls{cheops} sets a solid benchmark for future missions, offering valuable inspiration from the design and tailoring of the ground segment with a strong emphasis on the automation of activities.

\newpage
\appendix

\section{Hardware Architecture}

In this appendix, the hardware infrastructure hosting the \gls{cheops} subsystems performing Operations related tasks is briefly described. As the Mission and the Science Operations Centers are two physically distinct entities (near Madrid and Geneva respectively), they each have their own infrastructure, that are linked through an FTP server.

\subsection{Mission Operations Center}

\gls{moc} is composed of two platforms. The operational platform includes all the hardware and software used to actually monitor and control the spacecraft. Its architecture, shown in Figure\,\ref{fig:Test image}, is composed of two servers, three clients and a disk array where all the data are stored in RAID-5 configuration. Only the nominal server is able to read/write data on the array, but a change to the redundant configuration can be executed by switching off all tools, unmounting the disk from one server, mounting it in the other one and switching on the tools on the new server. The operational software is composed of the \gls{mcs}, \gls{fds}, automation system and file transfer tool. \gls{mcs}, \gls{fds} and the automation system are based on a server/client structure, with their server sides installed in all the servers and their client sides installed in all the clients. The file transfer tool is only installed in the servers. 

\begin{figure*}[h]
	\centering
	\includegraphics[width=\textwidth]{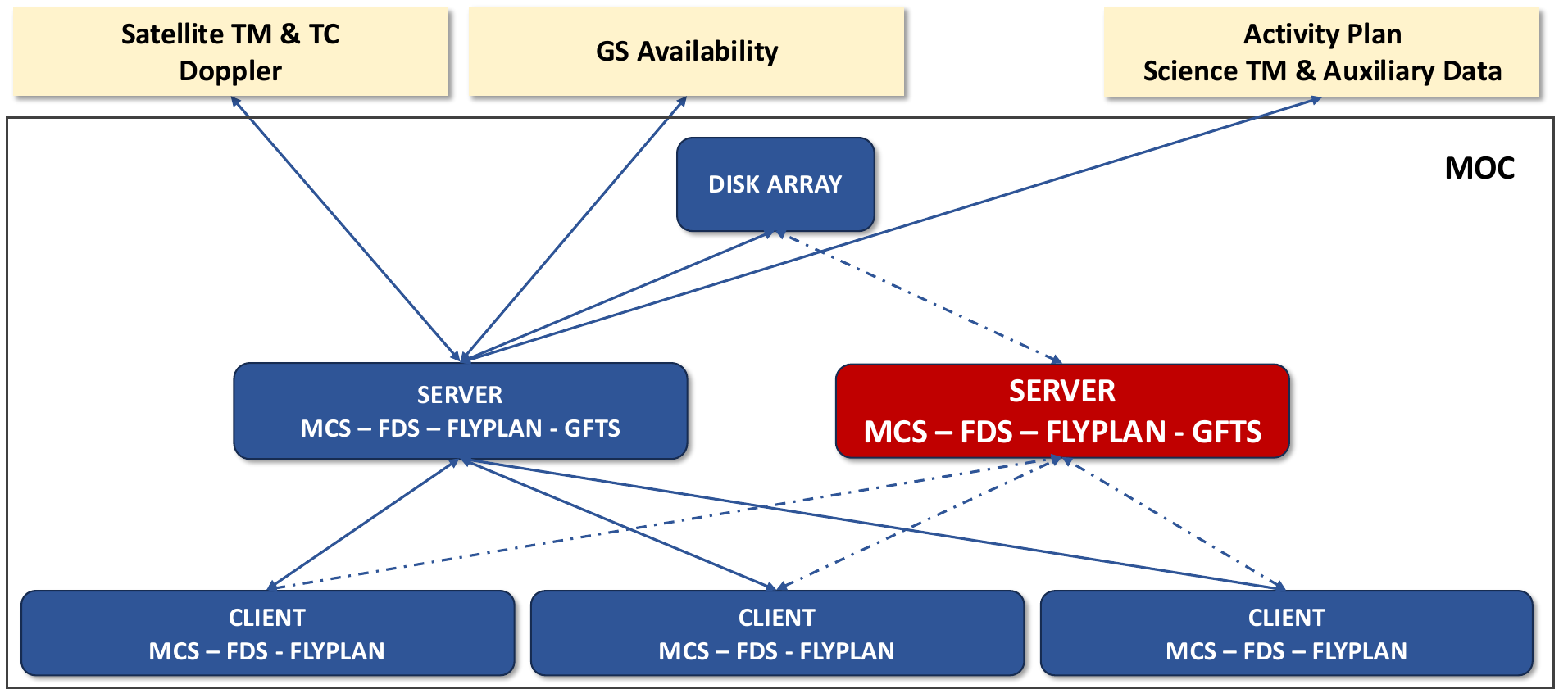}
	\caption{\gls{moc} Operational platform architecture. The solid arrows indicate the nominal configuration, whereas the dashed arrows illustrate the redundant configuration.}
	\label{fig:Test image}
\end{figure*}

The second platform is referred to as the reference platform and is used for training and validation. Originally based on a physical platform, it was migrated to a virtual environment in 2022 when hardware maintenance became problematic. As illustrated in Figure\,\ref{fig:Test image 2}, it is composed of two servers and one client. One of the servers and the client are configured in the same way as the ones in the operational platform (with the \gls{mcs}, \gls{fds} and automation servers) and the other server is used for the satellite simulator. The reference platform can be configured to access the \gls{gs} so it can be used as back up of the operational platform.

\begin{figure}[h! tbp]
	\centering
	\includegraphics[width=\columnwidth]{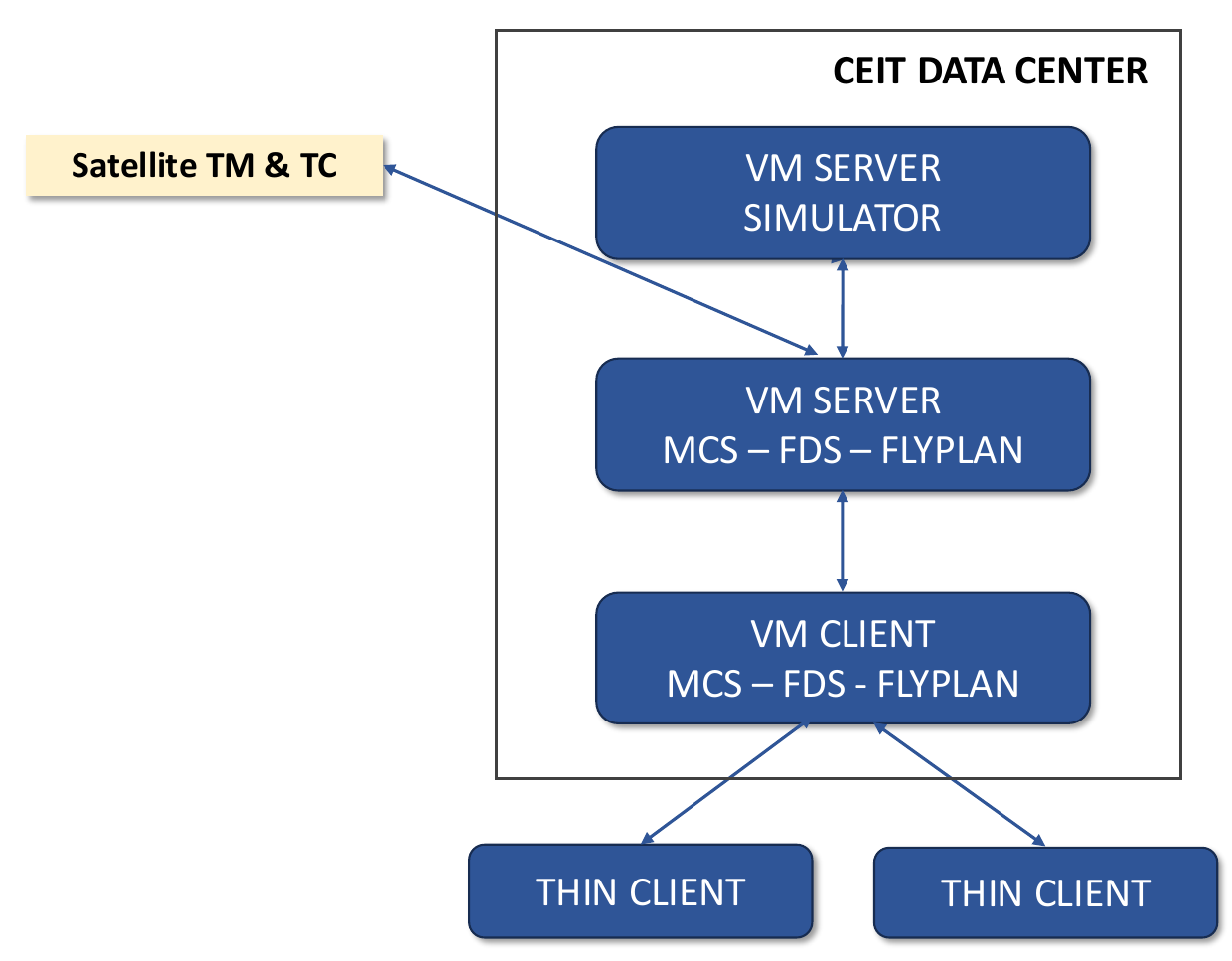}
	\caption{\gls{moc} Reference platform architecture.}
	\label{fig:Test image 2}
\end{figure}

\subsection{Science Operations Center}
\label{subsec:HWSOC}

The \gls{soc} system is based entirely on virtual machines and uses hardware infrastructure shared with other department-level projects for all of its processing and data storage needs. The system, which is shown in Figure\,\ref{fig:soc_computers}, consists of a web server for the \gls{pht2}, a server for the \gls{mps}, an FTP server for file transfer to and from \gls{moc}, a compute node running the data processing pipeline, with access to a computing cluster for resource intensive processing steps, a mission archive server and its accompanying web server hosting the archive web interface. The Space Science Data Center facility of the Italian Space Agency hosts mirrors of the archive and archive web servers.

\begin{figure}[h! tbp]
	\centering
	\includegraphics[width=\columnwidth]{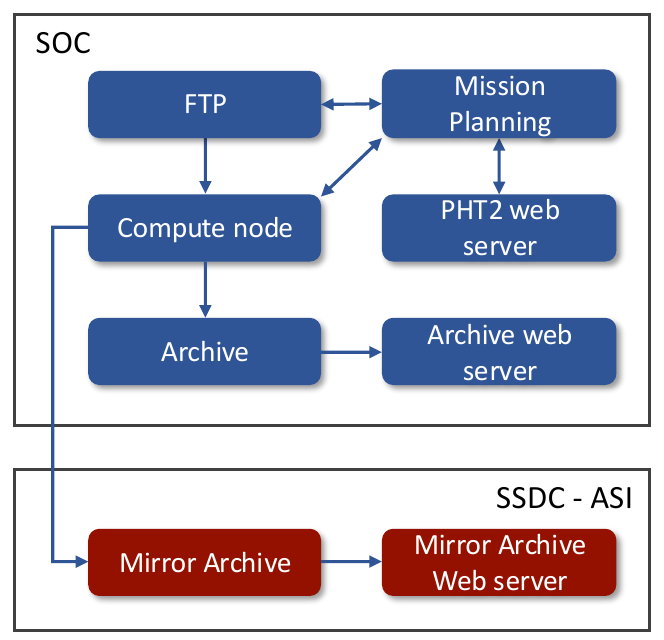}
	\caption{Overview of the hosts in the \gls{soc} system, the arrows denoting the flow of data through the system.}
	\label{fig:soc_computers}
\end{figure}

%\printglossary[type=\acronymtype, title=List of Acronyms, nonumberlist]
\printglossary[type=\acronymtype, title=List of Acronyms, style=myglossarystyle]

\printcredits

\section*{Acknowledgments}
\gls{cheops} is an ESA mission in partnership with Switzerland with important contributions to the payload and the ground segment from Austria, Belgium, France, Germany, Hungary, Italy, Portugal, Spain, Sweden, and the United Kingdom. The \gls{cheops} consortium would like to gratefully acknowledge the support received by all the agencies, offices, universities, and industries involved.

The Swiss ground segment development and operations were funded by the Swiss Space Office through the ANC and PRODEX programmes.

The Spanish \gls{moc} was funded by the Spanish Center for the Development of Technology and Innovation (CDTI) through the GSTP and PRODEX programmes.

After five years of operations, \gls{cheops} continues to run smoothly, an achievement made possible by the dedication and expertise of those who contributed to the development of its subsystems. The authors would like to express our heartfelt appreciation to the following individuals, whose flexibility, willingness to explore new approaches and invaluable efforts in the design, development, and management of the \gls{soc} have been instrumental to the success of \gls{cheops}: Mathias Beck, Reiner Rohlfs, David Futyan, 
Mohamed Meharga, Didier Queloz. We also wish to acknoledge all contributions from: Florian Georges, Magali Deleuil, Sergio Hoyer, Pascal Gutterman, Olivier Demangeon, Jean-Charles Meunier, Nick Walton, Guy Rixon, Gabor Kovács, Alexis Brandeker, Philip Loschl, Roland Ottensamer, Francesco Verrechia, Kate Isaak, Carlos Corral Van Damme and Willy Benz.

The authors also warmly acknowledge the contribution of the whole team at Deimos Engenharia, S.A.: Antonio Gutiérrez, Carlos António, Ignacio Garcia and Inês Estrela for the design and development of the \gls{mps} and the Archive, and their continuing support. Deimos Engenharia, S.A. also wishes to acknowledge contributions from past Deimos collaborators to the development of the \gls{mps}: Alejandro Martinez-Herrera, Diogo Andrade, Henrique Sousa, Marcos Bento, Paula Guerreiro, Pedro Neves, Rita Castro, Sofia Freitas, and Tural Malikli.

The authors express their gratitude to Richard Thomas Southworth for his invaluable contributions to mission operations, leveraging his extensive expertise at the European Space Operations Center. Additionally, they sincerely acknowledge José Ramiro Peñataro Blanco from GMV for his continuous technical support to daily operations. Their guidance has proven essential to the success of the operations throughout the mission. Finally and equally deserving of recognition, are the CEIT G/S and IT Network teams, whose indispensable contributions serve as the foundation of the MOC.

\section*{Declaration of generative AI and AI-assisted technologies in the writing process.}

During the preparation of this work, openAI's ChatGPT was used a few times by the authors to rephrase and clarify complex paragraphs. Each modification by ChatGPT was thoroughly reviewed to avoid any change in meaning and solely address clarity. The authors assume full responsibility for the content of the published article.

\bibliographystyle{cas-model2-names}

\bibliography{cas-refs}


\end{document}